\documentclass[english,conference]{IEEEtran}
\usepackage[T1]{fontenc}
\usepackage[latin9]{inputenc}
\usepackage{color}
\usepackage{amstext}
\usepackage{graphicx}
\usepackage{esint}

\makeatletter

\begin{document}
% paper title

\title{Calculation of the Autocorrelation Function of the Stochastic Single
Machine Infinite Bus System}

% author names and affiliations% use a multiple column layout for up to three different
% affiliations

\author{\authorblockN{Goodarz Ghanavati, Paul D. H. Hines and Taras Lakoba}
\authorblockA{College of Engineering and \\
Mathematical Sciences\\
 University of Vermont\\
 Burlington, VT\\
} \and 
\authorblockN{Eduardo Cotilla-Sanchez} \authorblockA{School
of Electrical Engineering\\
 and Computer Science\\
Oregon State University\\
Corvallis, OR}\\
}

% avoiding spaces at the end of the author lines is not a problem with
% conference papers because we don't use \thanks or \IEEEmembership

% for over three affiliations, or if they all won't fit within the width
% of the page, use this alternative format:
% 
%\author{\authorblockN{Michael Shell\authorrefmark{1},
%Homer Simpson\authorrefmark{2},
%James Kirk\authorrefmark{3}, 
%Montgomery Scott\authorrefmark{3} and
%Eldon Tyrell\authorrefmark{4}}
%\authorblockA{\authorrefmark{1}School of Electrical and Computer Engineering\\
%Georgia Institute of Technology,
%Atlanta, Georgia 30332--0250\\ Email: mshell@ece.gatech.edu}
%\authorblockA{\authorrefmark{2}Twentieth Century Fox, Springfield, USA\\
%Email: homer@thesimpsons.com}
%\authorblockA{\authorrefmark{3}Starfleet Academy, San Francisco, California 96678-2391\\
%Telephone: (800) 555--1212, Fax: (888) 555--1212}
%\authorblockA{\authorrefmark{4}Tyrell Inc., 123 Replicant Street, Los Angeles, California 90210--4321}}

% use only for invited papers
%\specialpapernotice{(Invited Paper)}

% make the title area
\maketitle
\begin{abstract}
Critical slowing down (CSD) is the phenomenon in which a system recovers more slowly from small perturbations. CSD, as evidenced by increasing signal variance and autocorrelation, has been observed in many dynamical systems approaching a critical transition, and thus can be a useful signal of proximity to transition. In this paper, we derive autocorrelation functions for the state variables of a stochastic single machine infinite bus system (SMIB). The results show that both autocorrelation and variance increase as this system approaches a saddle-node bifurcation. The autocorrelation functions help to explain why CSD can be used as an indicator of proximity to criticality in power systems revealing, for example, how nonlinearity in the SMIB system causes these signs to appear. 
\end{abstract}
% no keywords

% For peer review papers, you can put extra information on the cover
% page as needed:
% \begin{center} \bfseries EDICS Category: 3-BBND \end{center}
% for peerreview papers, inserts a page break and creates the second title.
% Will be ignored for other modes.
\IEEEpeerreviewmaketitle

\section{Introduction\label{sec:Introduction}}

Insufficient stability monitoring and situational awareness have been
identified as critical contributors to recent large power system failures,
such as August 14, 2003 \cite{August2003} and Sept.~8, 2011 \cite{Southwest}.
Recent deployment of phasor measurement units (PMU) produce large
quantities of time-series data, which open tremendous opportunities
to improve power system stability monitoring and control. However
doing so requires algorithms and methods that transform these data
into useful information about power system stability.

To address this problem, this paper draws from research on critical
transitions in stochastic dynamical systems. The statistics of time-series
data change notably in most stochastic dynamical systems as they approach
critical transitions; systems are more easily perturbed from equilibrium,
and take longer to return after being displaced \cite{scheffer2009early},
\cite{lenton2012early}. Collectively, this phenomenon is known as
Critical Slowing Down (CSD), and is most easily observed by testing
for autocorrelation and variance in time-series data. Increasing autocorrelation
and variance have been shown to indicate proximity to critical transitions
in climate models \cite{Slowclimate}, ecosystems~\cite{ecosystemslow}
and the human brain \cite{epilepticlitt2001}. In prior work by the
authors \textcolor{black}{\cite{CSDjournal}}, CSD has been shown
to be an early warning sign of critical transition in power systems.
A deep understanding of CSD in power systems could lead to new tools
for wide area measurement and control.

In many systems with critical transitions, slow-varying continuous
parameters cause gradual trends in state variables, which generally
move the system closer to, or further from, unstable operating points.
Simultaneously, random perturbations cause fast changes. As a result,
such systems have two time scales: fast and slow. In power systems,
loads have slow predictable trends, such as load ramps in the morning hours, and fast
stochastic ones, such as random load switching. Whereas conventional generators tend to vary slowly,
 intermittent renewable generation, such as PV on a partly cloudy day,
 can have fast stochastic changes. Because
of this, the mathematical framework of fast-slow systems \cite{kuehn2011mathematical}
and stochastic differential equations \cite{Hillstochastic} can help
to explain phenomena in these systems, such as the manner in which
autocorrelation and variance increase with proximity to critical points.
In~\cite{kuehn2011mathematical}, the variance and autocorrelation
of state variables for several prototypical fast-slow systems is calculated
using the Fokker-Planck approach and numerical simulations.

Estimating the proximity of a power system to a particular critical
transition (e.g., voltage collapse) has been the focus of a number
of papers in the power systems literature. References \cite{canizares2002voltage}\nocite{chiang1990voltage}\nocite{begovic1992control}-\cite{VanCutsem}
present methods to measure the distance between an operating condition
and voltage collapse with respect to slow-moving state variables,
such as load. While these methods provide useful information about
system stability, they require accurate network models all of which
contain some error.

Another approach to estimating the distance to critical transitions
is to identify statistical patterns in the response of a system to
stochastic forcing, such as fluctuations in load, or production from
renewable energy sources. To this end, a growing number of papers
study power system stability using stochastic models \cite{Bergen1987}\nocite{nwankpa1992stochastic,anghel2007stochastic,Hillstochastic}-\cite{Crowfokker}.
Reference \cite{Bergen1987} models power systems using Stochastic
Differential Equations (SDEs) and solves the SDEs using $\mathbf{\textrm{It\ensuremath{\hat{o}}}}$
calculus to develop a measure of voltage security. In \cite{Hillstochastic},
the Euler and Milstein methods for numerically solving SDEs are used
to assess transient stability in power systems, given fluctuating
loads and random faults. Reference \cite{Crowfokker} uses the time
evolution of the probability density function for state variables
in a Single Machine Infinite Bus (SMIB) system to show how random
load fluctuations affect system stability.

The results above clearly show that power system stability is affected
by noise in the system. However, more work is needed to identify useful
statistical trends in high sample-rate measurements from power systems.
Results from the literature on CSD suggest that the combination of
both increased autocorrelation and variance in time-series data are
needed in order to gain insight into the proximity to critical transitions
from data. Reference \textcolor{black}{\cite{CSDjournal} provides
empirical evidence of }increasing autocorrelation and variance for
an SMIB model and the 9-bus test case. Reference \cite{ChertkovCSD}
shows that voltage variance at the end of a distribution feeder increases
as it approaches voltage collapse. However, these results do not provide
insight into autocorrelation. To our knowledge, only \cite{Mitautocorrelation}
derives an approximate analytical autocorrelation function (from which
either autocorrelation or variance can be found) for state variables
in a power system model. However, the autocorrelation function in
\cite{Mitautocorrelation} is limited to the operating regime very
close to the threshold of instability. In this paper, we derive exact
autocorrelation functions for the state variables of an SMIB system.
The results provide insight into the conditions under which autocorrelation
and variance signal proximity to critical transitions in power systems.
\textcolor{black}{We use} the results to explain why CSD occurs in
power systems, and describe conditions under which autocorrelation
and variance signal proximity to critical transitions. 

The rest of the paper is organized as follows. Section \ref{sec:SDE}
provides a brief review of SDEs and solution methods. Section \ref{sec: SMIB}
describes our model, analytical and numerical results. Section \ref{sec:Discussion}
discusses the implications for power system operations, and Sec.~\ref{sec:Conclusion}
summarizes the results and contributions of this paper.

\section{A Brief Review of Stochastic Differential Equations\label{sec:SDE}}

There are two different approaches to modeling and solving Stochastic
Differential Equations (SDEs): the $\mathbf{\textrm{It\ensuremath{\hat{o}}}}$
and the Stratonovich interpretations. In the $\mathbf{\textrm{It\ensuremath{\hat{o}}}}$
interpretation \cite{ItoSDE}, noise is considered to be uncorrelated,
whereas in the Stratonovich interpretation \cite{Stratonovich:1963}
noise has finite, albeit very small, correlation time \cite{Gardiner:2004}.
$\mathbf{\textrm{It\ensuremath{\hat{o}}}}$ calculus is often used
in discrete systems, such as finance, though a few papers have applied
the $\mathbf{\textrm{It\ensuremath{\hat{o}}}}$ approach to power systems \cite{Bergen1987}, \cite{Hillstochastic}.
On the other hand, the Stratonovich method is often used in continuous
physical systems where noise is band-limited \cite{mannella2012ito}.
The Stratonovich interpretation also facilitates the use of ordinary
calculus, which is not possible under the $\mathbf{\textrm{It\ensuremath{\hat{o}}}}$
interpretation. 

In this paper we organize our model to use the multivariate Ornstein-Uhlenbeck
stochastic differential equation, as described in \cite{Gardiner:2004}:
\begin{equation}
d\underline{Z}\left(t\right)=A\underline{Z}\left(t\right)dt+Bd\mathbf{\underline{\textrm{W}}}\left(t\right)\label{eq:1-1}
\end{equation}
where $\underline{Z}\left(t\right)$ is the vector of the state variables,
$A$ and $B$ are constant matrices, and $\underline{W}\left(t\right)$
represents an $n$-dimensional Wiener process. The increments of the
Wiener process ($d\underline{W}(t)$) are independent, normally distributed
random variables, such that:
\begin{equation}
\underline{W}\left(t_{i}\right)-\underline{W}\left(t_{i-1}\right)\sim\mathcal{N}\left(0,t_{i}-t_{i-1}\right)\label{eq:4-1}
\end{equation}
Because $B$ is a constant matrix in this paper, the $\mathbf{\textrm{It\ensuremath{\hat{o}}}}$
and Stratonovich interpretations result in the same solution \cite{mannella2012ito}.
We will follow the Stratonovich interpretation because it allows for
the use of ordinary calculus.

\section{Stochastic Single Machine Infinite Bus System \label{sec: SMIB}}

Analysis of small power system models can be very helpful for understanding
the concepts of power system stability. The single machine infinite
bus system has long been used for this purpose. Reference \cite{demello1969concepts}
explores the small signal stability of synchronous machines using
the SMIB system. In \cite{wangHill1993SMIB}, Wang et al.~use a novel
control technique to improve the transient stability and voltage regulation
of a SMIB system. In the recent literature, there is increasing interest
in stochastic analysis of power systems, in part due to the increasing
integration of variable renewable energy sources. A few of these papers
use stochastic SMIB models. In \cite{wei2009SMIB}, it is suggested
that increasing noise in the stochastic SMIB system can make the system
unstable and induce chaotic behavior. Reference \cite{Crowfokker}
(mentioned in Sec.~\ref{sec:Introduction}) also studied stability
in a stochastic SMIB system.

In this section, we derive autocorrelation functions of the state
variables for a stochastic SMIB system. Analysis of these functions
provides analytical evidence for, and insight into, CSD in a small
power system.

\subsection{Stochastic SMIB System Model\label{sec: SMIB-1}}

\begin{figure}[h]

\includegraphics[width=1\columnwidth]{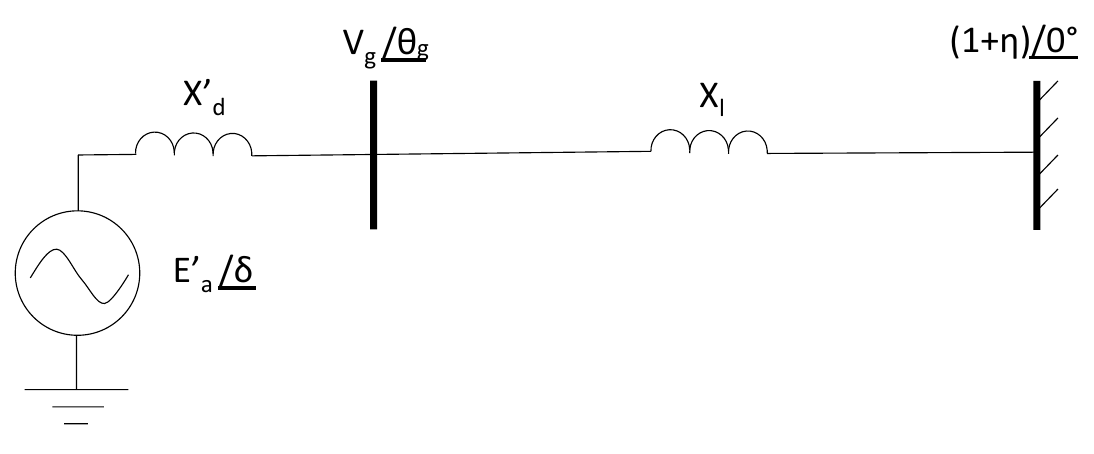}

\caption{\label{SMIB}Stochastic single machine infinite bus system model used
in sec. \ref{sec: SMIB}. }

\end{figure}

Fig. \ref{SMIB} shows an stochastic SMIB system. Equation (\ref{eq:1}),
which combines the mechanical swing equation and the electrical power
produced by the generator, fully describes the dynamics of this system:
\begin{equation}
M\ddot{\delta}+D\dot{\delta}+\frac{(1+\eta)E'_{a}}{X}\sin\left(\delta\right)=P_{m}\label{eq:1}
\end{equation}
where $(\eta\sim\mathcal{N}(0,0.01))$ is a white Gaussian random
variable added to the voltage magnitude of the infinite bus to account
for the noise in the system, $M$ and $D$ are the combined inertia
constant and damping coefficient of the generator and turbine, and
$E_{a}^{'}$ and $\delta$ are the transient emf and the rotor angle
of the generator. The rotor angle is the angle difference between
the rotor position and a synchronously rotating reference axis. The
reactance $X$ is the sum of the generator transient reactance $(X_{d}^{'})$
and the line reactance $\left(X_{l}\right)$, and $P_{m}$ is the
input mechanical power. The third term in the left-hand side of (\ref{eq:1})
is the generator's electrical power $\left(P_{g}\right)$. In order
to test the system with varying amounts of stress, we solved the system
for different equilibria, considering that generator's mechanical
and electrical power are equal at each equilibrium:
\begin{equation}
P_{m}=P_{g0}=\frac{E_{a}^{'}}{X}\sin\left(\delta_{0}\right)\label{eq:1-2}
\end{equation}

\subsection{Autocorrelation and Variance of the Differential Variables \label{sec: SMIB-2}}

In order to solve (\ref{eq:1}) analytically, we linearized it around
the equilibrium point using the first-order Taylor expansion:
\begin{equation}
\Delta\ddot{\delta}+\frac{D}{M}\Delta\dot{\delta}+\frac{E'_{a}}{MX}\cos\left(\delta_{0}\right)\Delta\delta=-\frac{\eta}{M}P_{g0}\label{eq:4}
\end{equation}
where $\delta_{0}$ is the value of the rotor angle at the equilibrium,
and $\Delta\delta$ is the deviation of the rotor angle from its mean
value. Equation (\ref{eq: damped ho}) is the standard form of (\ref{eq:4}),
which is known as the damped harmonic oscillator equation with noisy
forcing:
\begin{equation}
\ddot{Z}+2\gamma\dot{Z}+\omega_{0}^{2}Z=-f\eta\label{eq: damped ho}
\end{equation}
for which the following equalities hold:
\begin{equation}
Z=\Delta\delta;\gamma=\frac{D}{2M};\omega_{0}=\sqrt{\frac{E_{a}^{'}\cos\delta_{0}}{MX}};f=\frac{P_{g0}}{M}\label{eq:18}
\end{equation}
Equalities in (\ref{eq:18}) show that $f$ increases with $\delta_{0}$
while $\omega_{0}$ decreases with $\delta_{0}$.

Rewriting (\ref{eq: damped ho}) as (\ref{eq:1-1}) results in a set
of two first order SDEs. Here, $\underline{\textrm{Z}}\left(t\right)=\left[\begin{array}{cc}
\Delta\delta & \Delta\dot{\delta}\end{array}\right]^{T}$ where $\Delta\dot{\delta}$ is the deviation of the generator speed
from its mean value, and $\underline{W}(t)$ is a 2-variable Wiener
process. The matrices A and B in (\ref{eq:1-1}) are as follows:
\begin{equation}
A=\left[\begin{array}{cc}
0 & 1\\
-\omega_{0}^{2} & -2\gamma
\end{array}\right],\qquad B=\left[\begin{array}{cc}
0 & 0\\
0 & -f
\end{array}\right]\label{eq:7}
\end{equation}
At $\delta_{0}=\pi/2$, $\omega_{0}$ is equal to zero, so one of
the eigenvalues of matrix $A$ becomes zero. As a result, the system
experiences a saddle-node bifurcation. 

Following the method in \cite{Gardiner:2004}, the solution of (\ref{eq: damped ho})
is as follows:
\begin{eqnarray}
\Delta\delta\left(t\right) & = & f\cdot\int_{-\infty}^{t}[\exp\left(\gamma\left(t'-t\right)\right)\eta\left(t'\right)\label{eq:22-2}\\
 &  & \hspace{1em}\cdot\frac{\sin\left(\omega'(t'-t)\right)}{\omega'}]dt'\nonumber \\
\Delta\dot{\delta}\left(t\right) & = & f\cdot\int_{-\infty}^{t}[\exp\left(\gamma\left(t'-t\right)\right)\eta\left(t'\right)\label{eq:22-3}\\
 &  & \quad\cdot\frac{-\sin\left(\omega'(t'-t)+\phi\right)\omega_{0}}{\omega'}]dt'\nonumber 
\end{eqnarray}
where $\omega'=\sqrt{\omega_{0}^{2}-\gamma^{2}}$ is the frequency
of the underdamped harmonic oscillator, and $\phi=\arctan(\omega'/\gamma)$.
Note that $\omega'$ and $\phi$ decrease with $\delta_{0}$. Using
(\ref{eq:22-2}) and (\ref{eq:22-3}), we calculated the stationary
variances and autocorrelations of $\Delta\delta$, $\Delta\dot{\delta}$.
Note that the eigenvalues of $A$ have negative real part before the
bifurcation point since $\gamma>0$.   The variances are as follows:
\begin{eqnarray}
\sigma_{\Delta\delta}^{2} & = & \frac{f^{2}\sigma_{\eta}^{2}}{4\gamma\omega_{0}^{2}}\label{eq:8-1}\\
\sigma_{\Delta\dot{\delta}}^{2} & = & \frac{f^{2}\sigma_{\eta}^{2}}{4\gamma}\label{eq:8-2}
\end{eqnarray}
If $\gamma<\omega_{0}$, which holds until $\delta_{0}\approx\frac{\pi}{2}$
in our system, the autocorrelation functions for $\Delta\delta$ and
$\Delta\dot{\delta}$ are as follows:
\begin{eqnarray}
E\left[\Delta\delta\left(t\right)\Delta\delta\left(s\right)\right] & = & \exp\left(-\gamma\Delta t\right)\frac{f^{2}}{4\gamma\omega'\omega_{0}}\label{eq:10-1}\\
 &  & \cdot\sin\left(\omega'\Delta t+\phi\right)\sigma_{\eta}^{2}\nonumber \\
E\left[\Delta\dot{\delta}\left(t\right)\Delta\dot{\delta}\left(s\right)\right] & = & \exp\left(-\gamma\Delta t\right)\frac{-f^{2}\omega_{0}}{4\gamma\omega'}\label{eq:10-2}\\
 &  & \cdot\sin\left(\omega'\Delta t-\phi\right)\sigma_{\eta}^{2}\nonumber 
\end{eqnarray}
where $t$ and $s$ are two different times such that $t>s$, and
$\Delta t=t-s$.

\subsection{Numerical example \label{sub:SMIB-Numerical-example}}

\begin{figure}
\begin{centering}
\includegraphics[width=1\columnwidth,height=0.2\textheight]{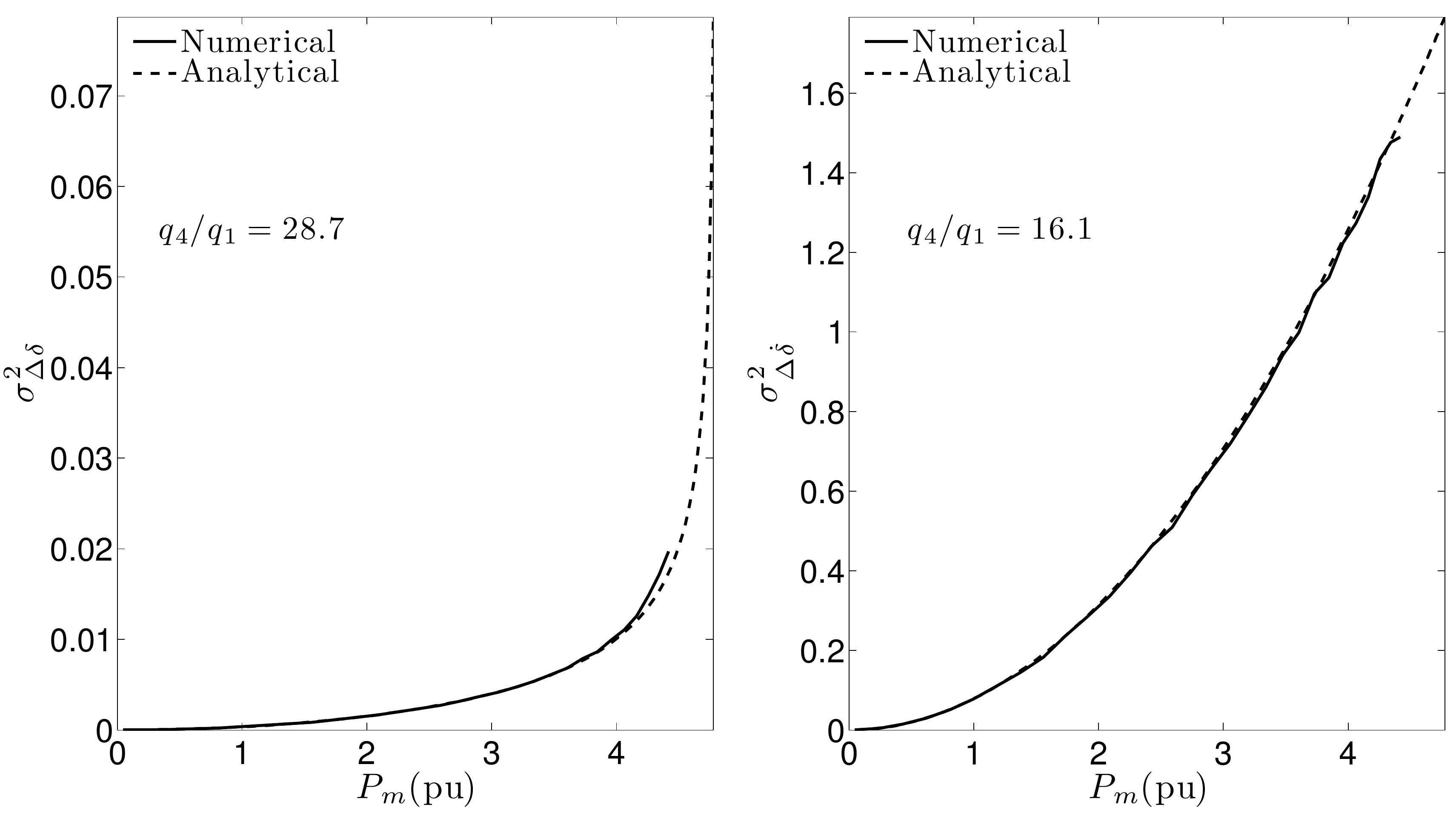}
\par\end{centering}

\caption{\label{fig_sys1_var_deom}Variance of $\Delta\delta,\Delta\dot{\delta}$
for different mechanical power $(P_{m})$ values. }
\vspace{-.1in}
\end{figure}

\begin{figure}
\begin{centering}
\includegraphics[width=1\columnwidth,height=0.2\textheight]{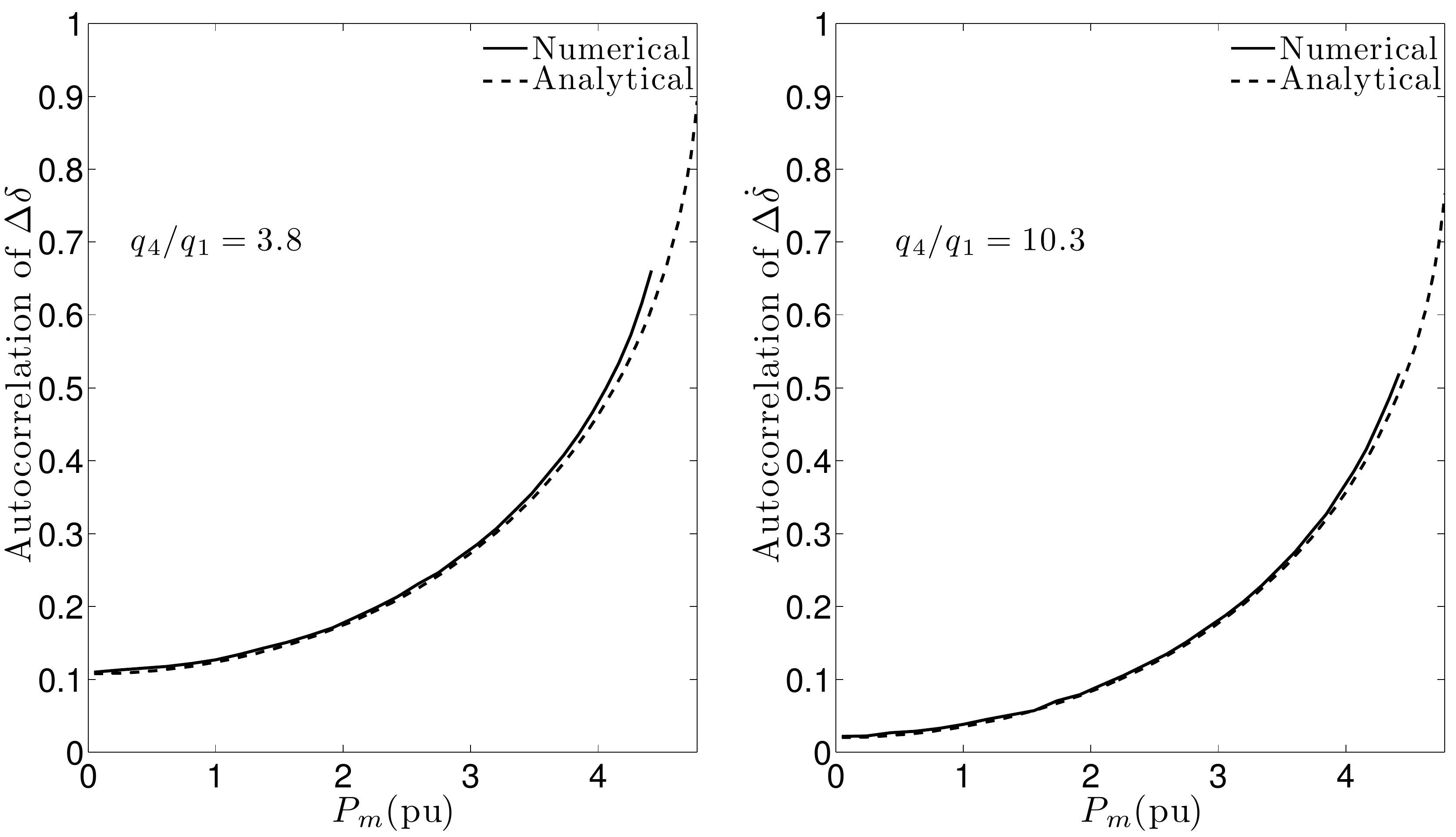}
\par\end{centering}

\caption{\label{fig_sys1_ar_deom}Autocorrelation of $\Delta\delta,\Delta\dot{\delta}$
for different mechanical power $(P_{m})$ values. The autocorrelation
values are normalized by dividing by the variances of the variables. }

\vspace{-.2in}
\end{figure}

Using (\ref{eq:8-1})-(\ref{eq:10-2}), we calculated the variances
and autocorrelations of $\Delta\delta$, $\Delta\dot{\delta}$ at
different equilibria; see Figs. \ref{fig_sys1_var_deom}, \ref{fig_sys1_ar_deom}.
Here, the bifurcation parameter is the input mechanical power $P_{m}$.
The parameters are given below: 

\begin{center}
$E'_{a}=1.2\textrm{pu},D=0.03\frac{\textrm{pu}}{rad/s},H=4\frac{MW.s}{MVA},X_{d}'=0.15\textrm{pu,}$
\par\end{center}

\begin{center}
$X_{l}=0.1\textrm{pu},\omega_{s}=2\pi\cdot60$
\par\end{center}

\noindent Note that $M=2H/\omega_{s}$, where $H$ is the inertia
constant, and $\omega_{s}$ is the rated speed in $rad/s$.

\begin{figure}
\begin{centering}
\includegraphics[width=1\columnwidth]{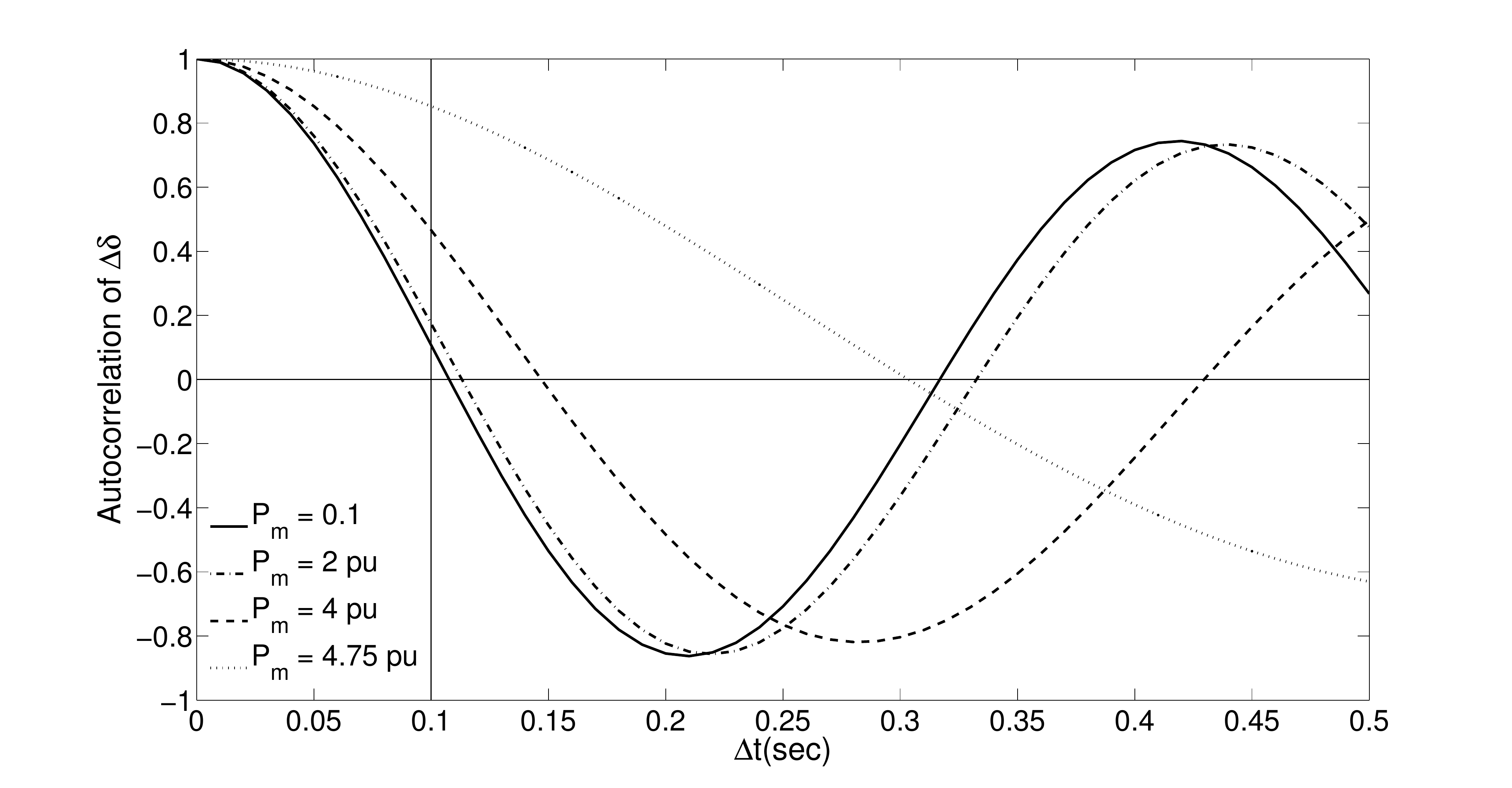}
\par\end{centering}

\caption{\label{fig: A_fcn}Autocorrelation function of $\Delta\delta$. $\Delta t=0.1$
is close to 1/4 of the smallest period of the function (for different
values of $P_{m}$). }
\vspace{-.3in}
\end{figure}

Fig. \ref{fig: A_fcn} shows the autocorrelation function of $\Delta\delta$
for different values of $P_{m}$. It shows that choosing $\Delta t$
close to 1/4 of the smallest period of the function  allows one
to observe the monotonic increase of the autocorrelation as $P_{m}$
increases. For larger values of $\Delta t$, the increase of the autocorrelation
may not be observed since the frequency of the function's oscillations
varies with $P_{m}$. For smaller values of $\Delta t$, the increase
of the autocorrelation is less noticeable since the function curves
become closer to each other as the time lag approaches zero. We chose
$\Delta t=0.1$s.  

In Figs. \ref{fig_sys1_var_deom} and \ref{fig_sys1_ar_deom}, the
analytical and numerical results are compared with each other. In
order to calculate the numerical results, (\ref{eq:1}) was solved
using a fixed-step trapezoidal ordinary differential equation solver.
At each time step, $\eta\left(t\right)$ changes according to its
normal probability density function. The minimum period of oscillations
in this system $(T=2\pi/\omega')$ is 0.4 sec. We chose the integration
step size to be 0.01 sec which is much shorter than the period of
the shortest oscillation. For each equilibrium, we integrated (\ref{eq: damped ho})
100 times; the average results are shown in the plots. The numerical
results are shown for the range of the bifurcation parameter values
for which the numerical solutions were stable. 
The ratio $q_{4}/q_{1}$ in Figs. \ref{fig_sys1_var_deom} and \ref{fig_sys1_ar_deom} is equal
to the value of the variance or autocorrelation for $P_{m}=4\textrm{pu}$ divided by the corresponding
 value for $P_{m}=1\textrm{pu}$.

Fig. \ref{fig_sys1_var_deom} shows that the variances of $\Delta\delta$
and $\Delta\dot{\delta}$ increase with $P_{m}$, and seem to be good
indicators of proximity to the bifurcation. However, the growth rates
of the two variances are different. The difference becomes more significant
near the bifurcation where the variance of $\Delta\delta$ increases
much faster than the variance of $\Delta\dot{\delta}$. This is caused
by the term $\omega_{0}^{2}$ in the denominator of the expression
for the variance of $\Delta\delta$ in (\ref{eq:8-1}) . In Fig. \ref{fig_sys1_ar_deom},
the autocorrelations of $\Delta\delta$ and $\Delta\dot{\delta}$
increase with $P_{m}$. Similar to the variances, the autocorrelations
are good indicators of proximity to the bifurcation as well. 

\subsection{Autocorrelation and Variance of the Algebraic Variables\label{sec: SMIB-3}}

In order to calculate the variance and autocorrelation of the algebraic
variables (the generator's terminal voltage magnitude and angle),
we wrote KCL at the generator's terminal:
\begin{equation}
\frac{E'_{a}e^{i\delta}-V_{g}e^{i\theta_{g}}}{jX'_{d}}+\frac{1+\eta-V_{g}e^{i\theta_{g}}}{jX_{l}}=0\label{eq:11-1}
\end{equation}
Separating the real and imaginary parts in (\ref{eq:11-1}), gives
the following:
\begin{eqnarray}
V_{g}\sin\left(\theta_{g}\right) & = & \alpha E_{a}^{'}\sin\left(\delta\right)\label{eq:11-2}\\
V_{g}\cos\left(\theta_{g}\right) & = & \alpha E_{a}^{'}\cos\left(\delta\right)\label{eq:11-3}\\
 &  & +\left(1+\eta\right)\left(1-\alpha\right)\nonumber 
\end{eqnarray}
where $\alpha=X_{l}/(X_{l}+X_{d}^{'})$. Linearizing (\ref{eq:11-2})
and (\ref{eq:11-3}) around the equilibrium results in $\Delta V_{g}$
and $\Delta\theta_{g}$ as linear combinations of $\Delta\delta$
and $\eta$:
\begin{eqnarray}
\Delta V_{g} & = & \alpha E_{a}^{'}\sin\left(\theta_{g0}-\delta_{0}\right)\Delta\delta\label{eq:13-1}\\
 &  & +\left(1-\alpha\right)\cos\left(\theta_{g0}\right)\eta\nonumber \\
\Delta\theta_{g} & = & (\alpha E_{a}^{'}\cos\left(\theta_{g0}-\delta_{0}\right)\Delta\delta\label{eq:13-2}\\
 &  & -\left(1-\alpha\right)\sin\left(\theta_{g0}\right)\eta)/V_{g0}\nonumber 
\end{eqnarray}
Then, we rewrote (\ref{eq:13-1}) and (\ref{eq:13-2}) as follows:
\begin{eqnarray}
\Delta V_{g} & = & C_{1}\Delta\delta+C_{2}\eta\label{eq:14-1}\\
\Delta\theta_{g} & = & C_{3}\Delta\delta+C_{4}\eta\label{eq:14-2}
\end{eqnarray}
where $C_{1},C_{2},C_{3},C_{4}$ are constants that replace the coefficients
in (\ref{eq:13-1}) and (\ref{eq:13-2}). Then, the autocorrelation
of $\Delta V_{g}$ is as follows $\left(t>s\right)$:
\begin{eqnarray}
E\left[\Delta V_{g}\left(t\right)\Delta V_{g}\left(s\right)\right] & = & C_{1}^{2}\cdot E\left[\Delta\delta\left(t\right)\Delta\delta\left(s\right)\right]\nonumber \\
 &  & +C_{1}C_{2}\mbox{\ensuremath{\cdot}}E\left[\Delta\delta\left(t\right)\eta\left(s\right)\right]\label{eq:15}
\end{eqnarray}
In deriving (\ref{eq:15}), we observed that $E\left[\Delta\delta\left(s\right)\eta\left(t\right)\right]=0$
since the system is causal. Also, $E\left[\eta\left(t\right)\eta\left(s\right)\right]=0$
because $\eta$ is a white random variable. Similarly,
\begin{eqnarray}
E\left[\Delta\theta_{g}\left(t\right)\Delta\theta_{g}\left(s\right)\right] & = & C_{3}^{2}\cdot E\left[\Delta\delta\left(t\right)\Delta\delta\left(s\right)\right]\label{eq:16}\\
 &  & +C_{3}C_{4}\cdot E\left[\Delta\delta\left(t\right)\eta\left(s\right)\right]\nonumber 
\end{eqnarray}
Equations (\ref{eq:15}), (\ref{eq:16}) show that in order to calculate
the autocorrelation of $\Delta V_{g}$ and $\Delta\theta_{g}$, it
is necessary to calculate $E\left[\Delta\delta\left(t\right)\eta\left(s\right)\right]$.
We calculated $E\left[\Delta\delta\left(t\right)\eta\left(s\right)\right]$
using (\ref{eq:22-2}):
\begin{eqnarray}
E\left[\Delta\delta\left(t\right)\eta\left(s\right)\right] & = & -\exp\left(-\gamma\Delta t\right)\cdot\frac{f}{\omega'}\label{eq:24-1}\\
 &  & \cdot\sin\left(\omega'\Delta t\right)\sigma_{\eta}^{2}\nonumber 
\end{eqnarray}
We can infer from (\ref{eq:24-1}) that $\textrm{cov}\left(\Delta\delta,\eta\right)=0$.
As a result, using (\ref{eq:8-1}), (\ref{eq:13-1}) and (\ref{eq:13-2}),
the variances of $\Delta V_{g}$ and $\Delta\theta_{g}$ are as follows:
\begin{eqnarray}
\sigma_{\Delta V_{g}}^{2} & = & \left(\frac{C_{1}^{2}f^{2}}{4\gamma\omega_{0}^{2}}+C_{2}^{2}\right)\sigma_{\eta}^{2}\label{eq:25-1}\\
\sigma_{\Delta\theta_{g}}^{2} & = & \left(\frac{C_{3}^{2}f^{2}}{4\gamma\omega_{0}^{2}}+C_{4}^{2}\right)\sigma_{\eta}^{2}\label{eq:25-2}
\end{eqnarray}
Utilizing (\ref{eq:10-1}), (\ref{eq:15}), (\ref{eq:16}) and (\ref{eq:24-1}),
we calculated the autocorrelation of $\Delta V_{g}$ :
\begin{eqnarray}
E\left[\Delta V_{g}\left(t\right)\Delta V_{g}\left(s\right)\right] & = & \exp\left(-\gamma\Delta t\right)\frac{C_{1}f}{4\omega'\omega_{0}\gamma}\cdot\label{eq:25-3}\\
 &  & \sqrt{C_{1}f\left(C_{1}f-8C_{2}\gamma^{2}\right)+\left(4C_{2}\omega_{0}\gamma\right)^{2}}\nonumber \\
 &  & \cdot\sin\left(\omega'\Delta t+\phi_{v_{g}}\right)\cdot\sigma_{\eta}^{2}\nonumber 
\end{eqnarray}
where $\phi_{v_{g}}=\arctan(\frac{C_{1}f\omega'}{C_{1}f\gamma-4C_{2}\gamma\omega_{0}^{2}})$.
The autocorrelation function of $\Delta\theta_{g}$ is similar to
(\ref{eq:25-3}) ($C_{1}$ and $C_{2}$ are replaced by $C_{3}$ and
$C_{4}$). 

Figs. \ref{fig_sys1_var_vgtg},\ref{fig_sys1_ar_vgtg} show the variance
and autocorrelation of $\Delta V_{g},\Delta\theta_{g}$ at different
equilibria for this system. In Fig. \ref{fig_sys1_var_vgtg}, although
the variances of both variables increase with $P_{m}$, the increase
of the variance of $\Delta\theta_{g}$ is much more significant. Also,
the variance of $\Delta V_{g}$ does not increase with $P_{m}$ until
the system gets close to the bifurcation while the variance of $\Delta\theta_{g}$
increases even if the system is far from the bifurcation. Both autocorrelations
of $\Delta V_{g}$ and $\Delta\theta_{g}$ in Fig. \ref{fig_sys1_ar_vgtg}
increase with $P_{m}$. However, the ratio $q_{4}/q_{1}$ is much
larger for $\Delta V_{g}$ than $\Delta\theta_{g}$.

\begin{center}
\begin{figure}[b]
\vspace{-.1in}
\begin{centering}
\includegraphics[width=1\columnwidth,height=0.2\textheight]{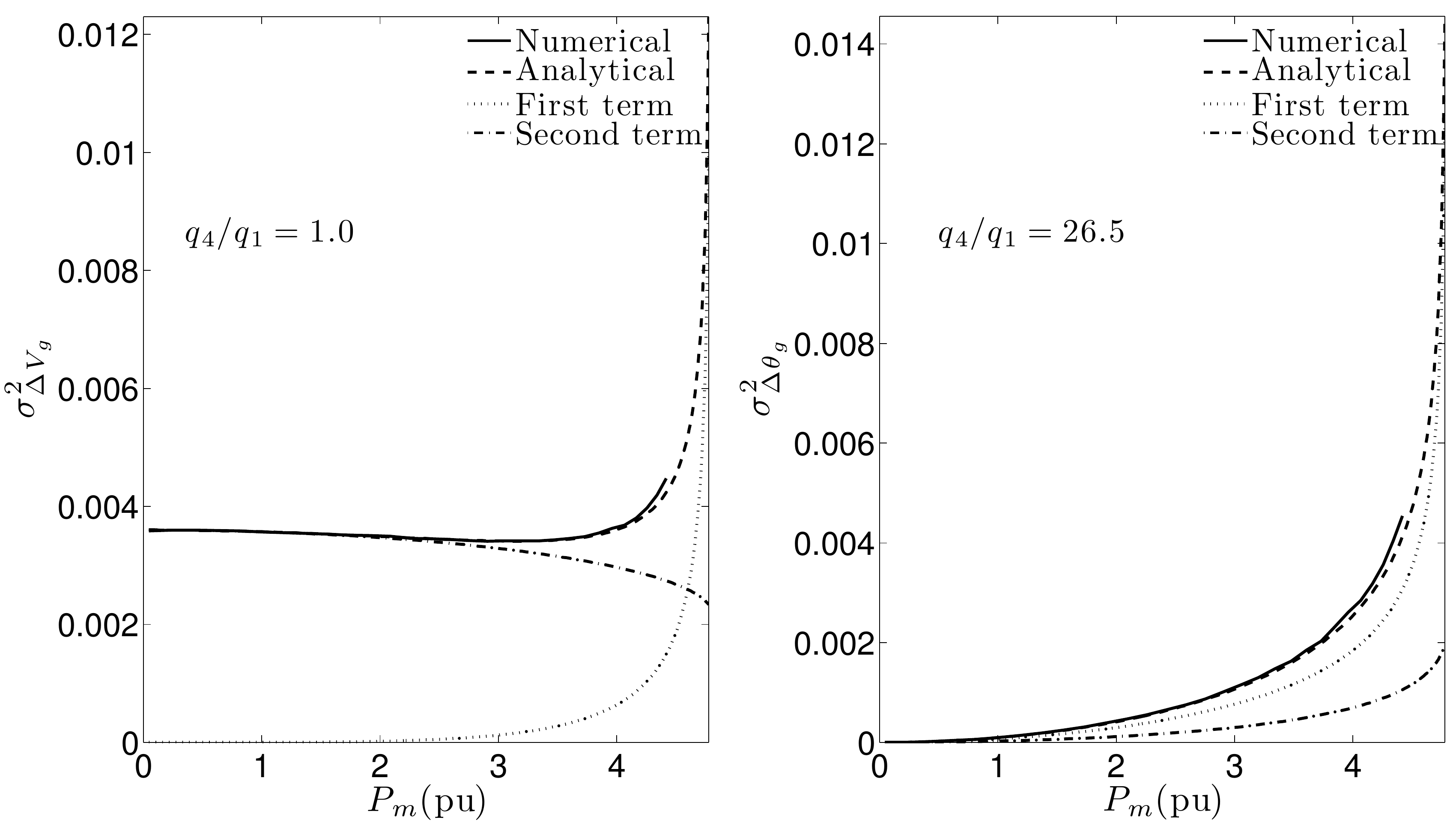}
\par\end{centering}

\caption{\label{fig_sys1_var_vgtg}Variance of $\Delta V_{g}$ and $\Delta\theta_{g}$
at different mechanical power $\left(P_{m}\right)$ levels. The two
terms comprising the variances in (\ref{eq:25-1}) and (\ref{eq:25-2})
are also shown.}
\vspace{-.1in}
\end{figure}

\par\end{center}

\begin{center}
\begin{figure}[t]
\begin{centering}
\includegraphics[width=1\columnwidth,height=0.2\textheight]{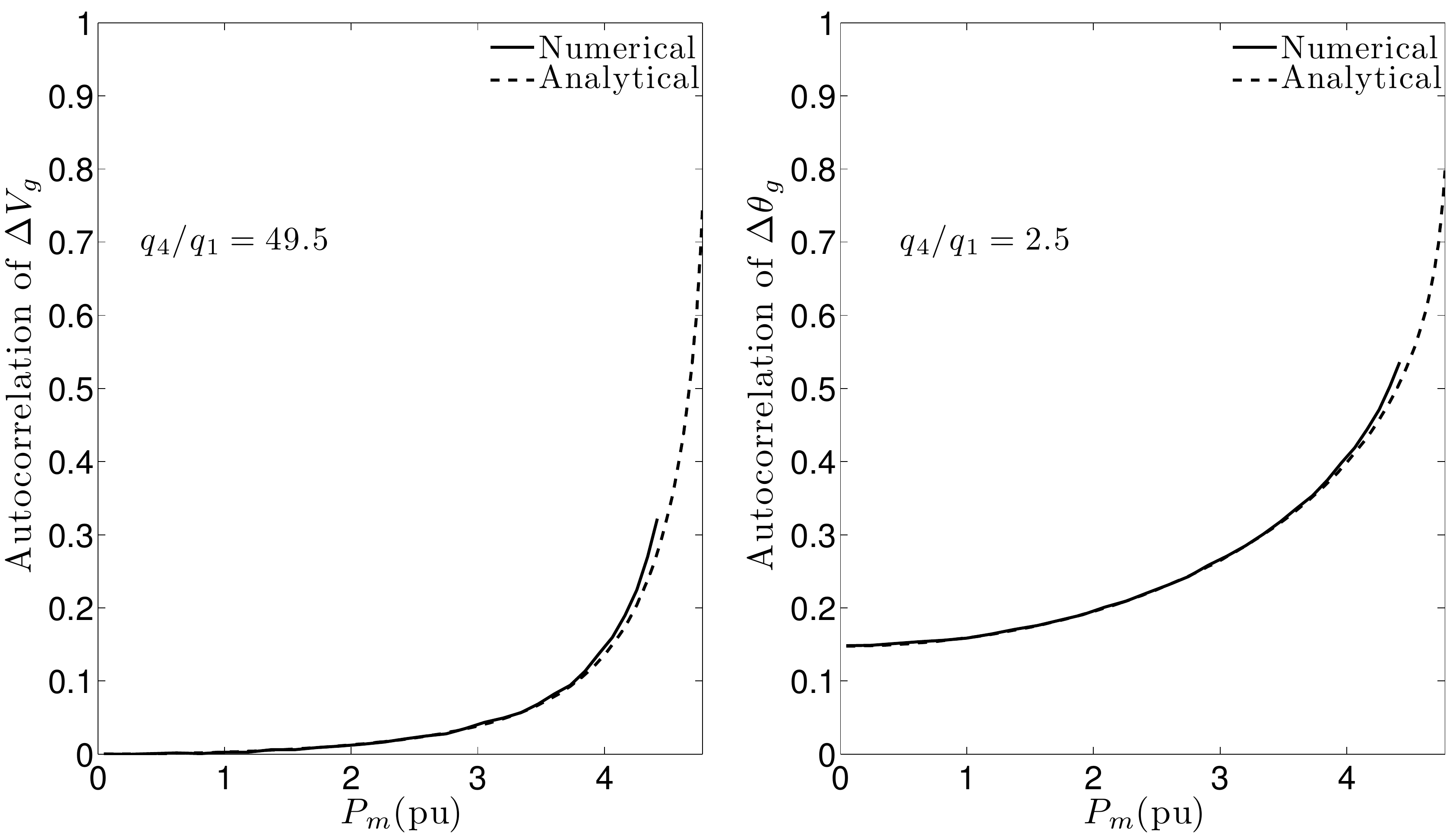}
\par\end{centering}

\caption{\label{fig_sys1_ar_vgtg}Autocorrelation of $\Delta V_{g}$ and $\Delta\theta_{g}$
versus $P_{m}$.}

\vspace{-.2in}
\end{figure}

\par\end{center}

\vspace{-.5in}

\section{Discussion \label{sec:Discussion}}

\begin{figure}[b]
\vspace{-.2in}
\includegraphics[width=1\columnwidth]{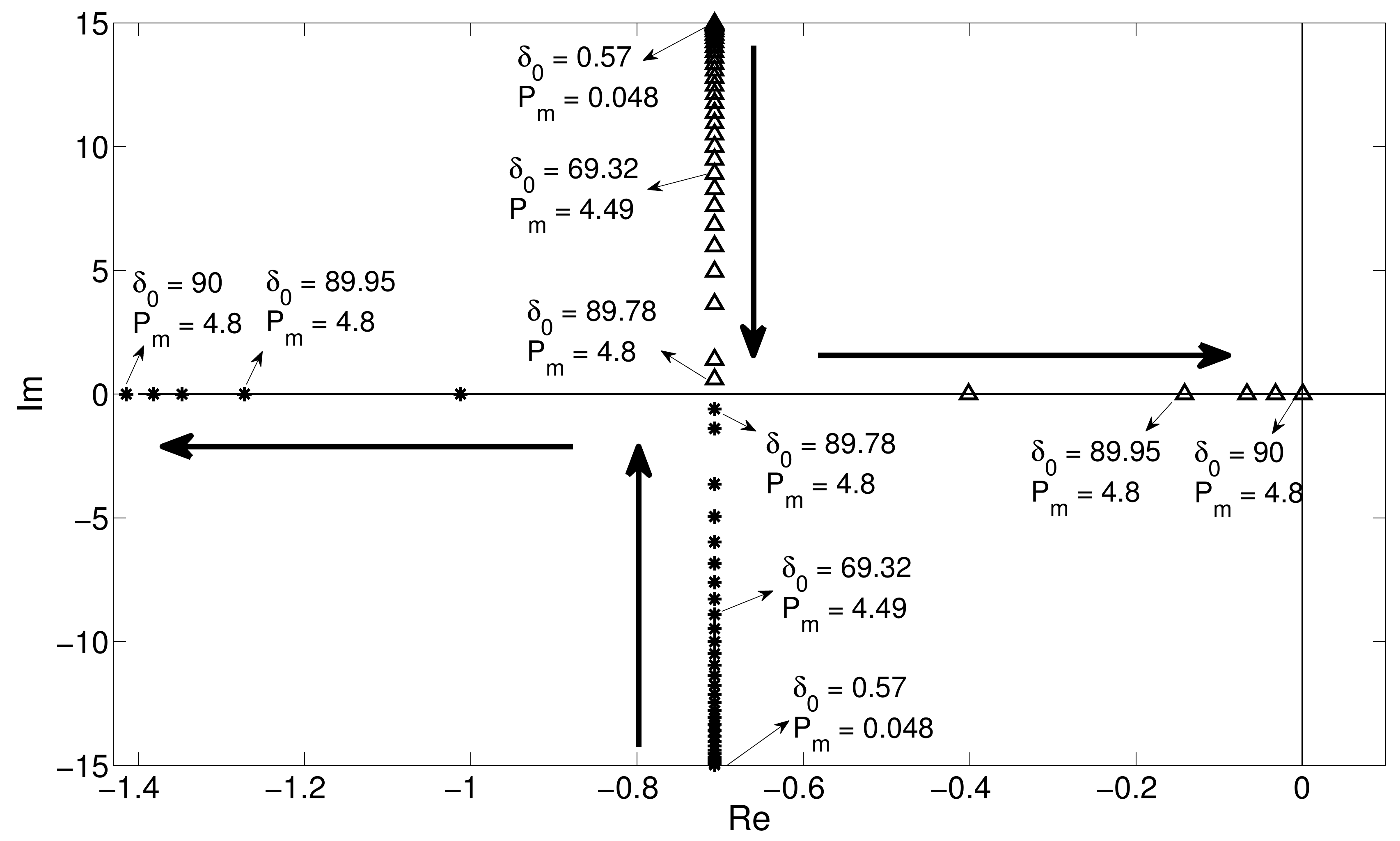}

\caption{\label{sys1_eigs}Eigenvalues of the first system as the bifurcation
parameter (mechanical power) is increased. The arrows show the direction
of the eigenvalues' movement in the complex plane as $P_{m}$ is increased.
The values of $P_{m}$ and $\delta_{0}$ are given for several eigenvalues.}

\vspace{-.3in}
\end{figure}

Fig.~\ref{sys1_eigs} shows the trajectory of the eigenvalues of
the state matrix $A$. The system passes through a saddle-node bifurcation
as the mechanical power is increased. Near the bifurcation, the eigenvalues 
are very sensitive to the change of the bifurcation parameter. As a result, the 
system is in the overdamped regime $(\omega_{0}<\gamma)$
  for much less than 0.1\% distance in terms of
$P_{m}$ to the critical transition. This is consistent with \cite{Mitautocorrelation},
 where the autocorrelation function is valid when the system is within 0.1\%
  to the saddle-node bifurcation such that the lowest eigenvalue determines the 
system dynamics. Accordingly, it can provide a good estimate of the autocorrelation
 and variance of the system states for a very short range of the bifurcation parameter,
 but the result of \cite{Mitautocorrelation} can not be used as an early warning sign.

Figs.~\ref{fig_sys1_var_deom}-\ref{fig_sys1_ar_vgtg} show that
the variance and autocorrelation of all four state variables increase
when the system is more stressed. This demonstrates that CSD occurs
in this system as it approaches the bifurcation, as suggested both
general results \cite{kuehn2011mathematical}, and prior work for
power systems \cite{CSDjournal,Mitautocorrelation}. 

In addition to validating these prior results, several new observations
can be made. For example, the signs of CSD are more clearly observable
in some variables, and not others. While all of the variables show
some increases in autocorrelation and variance, they are less clearly
observable in $\Delta V_{g}$. The variance of $\Delta V_{g}$ decreases
with $P_{m}$ slightly until the vicinity of the bifurcation. On the
other hand, the variance of $\Delta\theta_{g}$ always increases with
$P_{m}$. Fig.~\ref{fig_sys1_var_vgtg} shows the two terms in the
expressions for $\sigma_{\Delta V_{g}}^{2}$ and $\sigma_{\Delta\theta_{g}}^{2}$,
from (\ref{eq:25-1}) and (\ref{eq:25-2}). The first term in $\sigma_{\Delta V_{g}}^{2}$
is always dominant and growing. On the other hand, the second term
in $\sigma_{\Delta V_{g}}^{2}$ is larger for small $P_{m}$. The fact that
this term decreases with $P_{m}$ can be observed from $C_{2}$ in
(\ref{eq:13-1}). Accordingly, the decrease of $C_{2}$ with $P_{m}$
causes $\sigma_{\Delta V_{g}}^{2}$ to decrease until $P_{m}$ reaches
the vicinity of the bifurcation. With respect to $\sigma_{\Delta\theta_{g}}^{2}$,
the second term in (\ref{eq:25-2}) increases with $P_{m}$, and adds
to the increase caused by the first term. In conclusion, the variance
of $\Delta\theta_{g}$ is a better indicator of proximity to the bifurcation. 

The results also show that the nonlinearity of this system causes
CSD to occur within it. In (\ref{eq:7}), one of the elements of the
state matrix $(-\omega_{0}^{2})$ changes with $P_{m}$ because of
the nonlinear relationship between the electrical power $(P_{g})$
and the rotor angle, and that element causes the eigenvalues to change
with $P_{m}$. If the relationship between $P_{g}$ and $\delta$
were linear, the state matrix would be constant. \textcolor{black}{In
\cite{Gardiner:2004}, it is shown that the }stationary time correlation
matrix \textcolor{black}{of (\ref{eq:1-1})} can be calculated using
the following equation\textcolor{black}{:}
\begin{equation}
E\left[\mathbf{\mathbf{\underline{\mathrm{\mathit{Z}}}\left(\mathit{t}\right)}}\underline{Z}{}^{T}\left(s\right)\right]=\exp\left[-A\Delta t\right]\sigma\label{eq:26}
\end{equation}
where $\sigma$ is the covariance matrix of the state variables.\textcolor{black}{{}
}From (\ref{eq:26}), it can be inferred that t\textcolor{black}{he
normalized autocorrelation matrix only depends on $A$ and the time
lag}. As a result, if the state matrix is constant, the autocorrelations
for an specific time lag will be constant as well. Accordingly, in
this system, CSD is caused by the nonlinear relationship between $P_{g}$
and the rotor angle.

Further research is required to design methods that can use CSD to accurately determine whether a particular state is approaching instability in larger power systems. However, the following are some ideas for using this method in a real power system. 
Using the analytical approaches in this paper in combination with data, we expect to be able to estimate baseline (normal) values for the variance and autocorrelation of system variables. Doing so will require some extension of the analytical approaches in this paper. While it is unlikely that this approach can be used to produce explicit autocorrelation functions, the method should be able to produce numerical values for autocorrelations and variances. Also, using the approach outlined here, the ratio of the statistics in critical conditions to those in normal 
conditions can be calculated from off-line studies. Using the base values and critical ratios, one can calculate thresholds for CSD signs, which help in informing system operators regarding a critical condition.
Calculating autocorrelation and variance from data is not computationally expensive, which makes this method well-suited for real-time application in large power grids.

\section{Conclusion\label{sec:Conclusion}}

In this paper, we derive analytical autocorrelation functions for the state variables in a stochastic single machine infinite bus system. The analytical results were confirmed by numerical simulations. The functions show that this system does indeed express critical slowing down, as evidenced by both increased autocorrelation and variance, before the bifurcation occurs. We also showed that the occurrence of CSD in this system is due to its nonlinearity. Moreover, the results reveal that there are some differences in the growth rate of the CSD signs for different state variables; some variables are better indicators of proximity to the bifurcation than others. These findings suggest ways to better understand CSD in large multi- machine power systems. Understanding the statistical behavior of stochastic power systems as they approach instability should allow for the development of new indicators of power system stability based on the statistical properties of PMU data.

% conference papers do not normally have an appendix

% use section* for acknowledgement

\section*{Acknowledgment}

The authors acknowledge the Vermont Advanced Computing Core, which
is supported by NASA (NNX 06AC88G), at the University of Vermont for
providing High Performance Computing resources that have contributed
to the research results reported in this paper.

% optional entry into table of contents (if used)

% trigger a \newpage just before the given reference
% number - used to balance the columns on the last page
% adjust value as needed - may need to be readjusted if
% the document is modified later
%\IEEEtriggeratref{8}
% The "triggered" command can be changed if desired:
%\IEEEtriggercmd{\enlargethispage{-5in}}

% references section
% NOTE: BibTeX documentation can be easily obtained at:
% http://www.ctan.org/tex-archive/biblio/bibtex/contrib/doc/

% can use a bibliography generated by BibTeX as a .bbl file
% standard IEEE bibliography style from:
% http://www.ctan.org/tex-archive/macros/latex/contrib/supported/IEEEtran/bibtex
%\bibliographystyle{IEEEtran.bst}
% argument is your BibTeX string definitions and bibliography database(s)
%\bibliography{IEEEabrv,../bib/paper}
% <OR> manually copy in the resultant .bbl file
% set second argument of \begin to the number of references
% (used to reserve space for the reference number labels box)

% Generated by IEEEtran.bst, version: 1.13 (2008/09/30)

\end{document}